\documentclass[fleqn,twoside]{article}
\usepackage{espcrc2}

\usepackage{graphicx}
\usepackage[figuresright]{rotating}

\newcommand{\AmS}{{\protect\the\textfont2
  A\kern-.1667em\lower.5ex\hbox{M}\kern-.125emS}}

\begin{document}

\title{Centauros and/or Chirons as evaporating mini black holes}

\author{Theodore N. Tomaras\address{Department of Physics and Institute of
Plasma Physics, University of Crete, \\
P.O.Box 2208, 71003 Heraklion, Crete, Hellas \\ 
and Fo.R.T.H.} 
        \thanks{This talk is mostly based on \cite{mmt}. 
	Research supported in part by the EU under the RTN contract
	HPRN-CT-2000-00122, and by a grant P.O. Education - Heraklitos
	from the Hellenic Ministry of Education.}}
       

\begin{abstract}
It is argued that the signals expected from the evaporation of mini 
black holes - predicted in TeV-scale gravity models with large extra 
dimensions, and possibly produced in ultra high energy collisions in the
atmosphere - have characteristics quite similar to the ones of the Centauro
events, an old mystery of cosmic-ray physics.

\vspace{1pc}
\end{abstract}

\maketitle

\section{Introduction}

The theoretical framework of this talk is the {\it TeV-scale gravity
with large extra dimensions}. The basic assumption, easily accomodated in 
Superstring Theory, is that our spacetime is a 4 dimensional hypersurface
(a {\it D3-brane}) embedded in a 10 dimensional world (the {\it bulk}), 
with the additional feature
that all known matter (quarks, leptons, higgses), as well as the carriers of
the fundamental interactions of the Standard Model of Particle Physics 
(photon, W, Z, gluons) are
confined on our 4 dimensional subspace. Only gravity can propagate in
the bulk. The fact that the only communication with the bulk is via the 
gravitational force, allows for a fundamental gravitational scale to be
of ${\cal O}$(1TeV), while at the same time the size of the extra spatial
dimensions may be as large as a fraction of a mm. 

Clearly, such a scenario, if true, leads to the exciting prospect of 
observing string physics, large higher dimensions and quantum gravity effects, 
within the next few years in the forthcoming accelerator (LHC), neutrino and 
cosmic-ray experiments.

Even better, in this talk I will argue that the long known Centauro-like 
events (CLEs) may be due to the formation and subsequent evaporation of 
mini black holes
(MBHs), predicted in the context of the above scenario \cite{mmt}. 
After a quick review of the relevant scales and of the main features of 
MBHs in TeV-gravity, as well as of the basic characteristics of the 
Centauro events, I present in Section 5 the somewhat qualitative arguments
that support our interpretation. The talk ends with a critical discussion.

The first steps towards a more quantitative analytical/Monte Carlo 
investigation of the mini black 
hole picture are described in \cite{cct} and \cite{cct2}.

\section{TeV-scale gravity models}

Even though non-compact extra dimensions are not a priori excluded,
the most straightforward realization of the above theoretical scenario 
is to neglect the
tension on the brane, which would modify the gravitational background, 
and to consider
the extra dimensions compact and forming a higher dimensional torus $T^n$, 
with equal radii $R$.  
 
In the simplest case of $n=1$, one is dealing with a 4 dimensional Minkowski
space with one extra transverse dimension, a circle of radius $R$. 
One may compute the gravitational potential $\Phi({{\bf x},\theta})$
(in the Newtonian approximation) of a point mass $M$ at the origin, by solving
the corresponding Poisson equation
\begin{equation}
\nabla^2 \Phi+{1\over R^2}\frac{\partial^2\Phi}{\partial\theta^2}=
-G^{(4)}M\delta({\bf x})\frac{\delta(\theta)}{2\pi R}
\label{poisson}
\end{equation}
where $G^{(4)}\equiv 1/M_*^3$ is the 4 dimensional gravitational constant,
with fundamental gravitational scale equal to $M_*$.
Its solution is 
\begin{equation}
\Phi({\bf x},\theta)=\frac{2\pi M G^{(4)}}{Rr}
\frac{1-e^{-2r/R}}{1-2e^{-r/R}\cos\theta+e^{-2r/R}}
\label{potential}
\end{equation}
where $r=|{\bf x}|$ and $\theta$ is the position in the compact
transverse direction. 
Near the mass $M$, i.e. for $r\ll R$ and $\theta\ll\pi$ it gives 
$\Phi\sim 4\pi MG^{(4)}/(r^2+R^2\theta^2)$, the correct expression for a 
4 dimensional gravitational potential. At large distances ($r\gg R$) 
from the mass $M$, on the other hand, one obtains 
$\Phi\sim 2\pi M G^{(4)}/Rr$, the correct 3 dimensional Newtonian 
potential with Newton's constant $G_N=2\pi G^{(4)}/R=2\pi/RM_*^3$. 

These formulas generalize trivially to n toroidal 
transverse dimensions, in which case
Newton's constant is given by
\begin{equation}
G_N\sim {1\over {M_*^2}}{1\over{(RM_*)^n}}
\label{GN}
\end{equation}

{\it Assume} that the fundamental scale $M_*$ of gravity is of the order of the weak 
interaction scale $M_*\sim {\cal O}$(1TeV). Then, (\ref{GN}) leads to the 
correct value $G_N\simeq 10^{-38}$GeV$^{-2}$ for a value of R given by
$R_n\simeq 1/(M_*(M_*^2 G_N)^{1/n})$. 
The value $n=1$ is excluded, since it leads to R of the order of the size 
of the solar system. For $n=2$ one obtains $R_2\sim 2$mm. 
Accelerator and astrophysical constraints lead to a prefered value of $n\geq 4$
with a corresponding value for R, much larger than the usual value $10^{-33}$cm.

\section{Mini black holes}

As described, the world contains black holes, generalizations to $D=4+n$
spacetime dimensions of the well known Schwarzschild metric.
They are characterized by their mass $M_{BH}$, and can exist as long as
$M_{BH}$ is at least a few times the fundamental gravity scale. Lighter black holes 
cannot exist. Thus, it is reasonable to expect that for $M_*$ a fraction of a TeV, 
the black hole masses are $M_{BH}\geq 2$TeV. 
If, in addition, 
$M_{BH}\ll M_* (G_N M_*^2)^{-(n+1)/n}$, the black hole is essentially 4+n-dimensional,
since its Schwarzschild radius $R_S^{-1}\sim M_* (M_*/M_{BH})^{(1/n+1)}\sim M_*$ 
is much smaller than the radius of the large extra dimensions.
Black holes are expected to be produced in the collision of any two particles, as long
as their impact parameter is smaller than the Schwarzschild radius corresponding
to their center of mass energy. A black hole with mass of the order of the ones
discussed here can be produced in the collision of an ultra high energy primary with
atmospheric partons. The energy of the primary should exceed 1000 TeV and the 
cross section of the process is conjectured to be $\sigma\sim \pi R_S^2 \sim 10^{-37}cm^2
\sim \sigma_{\nu N}^{weak}$, comparable to the neutrino-nucleon
weak interaction cross section at these energies.  

Once produced, black holes are believed to evaporate. 
Even though noone has worked out the details of the evaporation process for such
light black holes, we shall assume the semiclassical formulas of the standard 
treatment. So, within $\tau_{BH}\sim M_*^{n+2} (4\pi R_S)^{n+3}/(n+1)^{n+3}\sim
10^{-27}$seconds the black hole decays "democratically" into all kinds of 
quarks, leptons, gauge bosons, gravitons, higgses. It is a fireball of temperature
$T_{BH}=(n+1)/4\pi R_S\sim 1$TeV. The number of initial particles emitted
by the black hole is determined by its entropy $N_{\rm initial}
\sim S_{BH}=\pi M_{BH}R_S/2$. For the case of black holes with mass of order 1-2 TeV 
of interest here and $n=4$, this number is of the order of ${\cal O}$(10).

\section{The Centauro-like events (CLEs)}

A normal high energy cosmic ray event is created by the collision of a primary
particle with a particle in the atmosphere. Typically a couple of leading partons
emerge from the interaction region with high transverse momenta, leaving
behind a number of soft fragments, mainly pions with relative abundance 1:1:1. 
The neutral pions subsequently decay to photons and the shower ends up
consisting of low $p_T<1$ GeV/c particles with $N_{\rm hadron}/N_{\rm em}\sim 1$ 
or even less, if one takes into account the extra photons that will be produced
as the shower develops even further.

In contrast to the above picture, several events have been observed since 1972
with the following main characteristics \cite{ewa}:
(a) They were claimed to have taken place at distances smaller that 500 m 
above the detectors at Pamir and Chacaltaya. It should be pointed out, however, 
that the altitude was measured directly only for one of these events, and even that
has been questioned recently \cite{OST}. (b) They are hadron rich, with typically
$N_{\rm hadron}/N_{\rm em}\gg 1$, (c) have fragments with high $p_T\gg 1$ GeV/c,
(d) have in many cases a heavy central core with tiny angular opening (halo) and finally, 
(e) have all been observed with energies above a threshold around 
500 TeV in the lab frame.

It should be pointed out that there are severe uncertainties in the observational
data. One of the speakers in this meeting presented a reanalysis of the Centauro I 
and raised serious doubts about the altitudes reported in general 
for all these events \cite{OST}. Others express doubts about the existence of these
events altogether, worrying, for instance, 
about the fact that no such events have been observed
yet in Kanbala and Fuji. We shall not take part in this debate at this point. 
Instead, we shall assume that the events are real and try to interpret them 
as due to evaporating MBHs, produced by ultra high energy 
$E_1>1000$TeV cosmic ray primaries.

\section{CLEs as evaporating mini black holes}

The processes of black hole creation, of its subsequent  evaporation and, finally
of the shower formation in the atmosphere involve all the complications
of several different fields, such as cosmic ray physics, quantum gravity/string theory,
quantum field theory in curved spacetimes,
quark-gluon plasma physics in QCD, low energy non-perturbative QCD,
cosmic shower atmospheric physics, of theoretical and experimental high energy
physics and astrophysics. Given the incomplete knowledge in all these,
a lot has to be done before one can safely confront the observational data.
Nevertheless, we shall make a few simplifying assumptions, implicit in the 
discussion below, and present the arguments 
in favour of the scenario proposed in \cite{mmt,cct,cct2}.

$\bullet$ {\it Energy threshold}. The energy threshold of the observed Centauro
events, corresponds in the center of mass frame to a mass roughly a few times 
$M_*\sim 1$TeV. This coincides with the lower bound on the MBH masses,
mentioned above. The agreement may be even better, if one takes into account 
the energy losses into the bulk 
during the evaporation of the MBH. 

$\bullet$ {\it Production rate}. Assuming that the black holes are produced by
primary neutrinos, one may obtain a rough estimate of the number of
CLEs based on current figures for the neutrino flux $\Phi_\nu$. 
Since we are interested in neutrino energies of order $10^6-10^7$GeV in the lab frame,
one may use the estimates for the gamma-ray burst muon neutrino flux given in
\cite{wb,w}. Their analysis leads to $\sim$20 neutrino-induced muon events in a 
km$^3$ water or ice per year. Since the cross section of MBH production
by neutrinos is of the same order of magnitude as that of muon production, we would
expect approximately the above number of black holes for each kind of neutrino. 
Multiplying by 10-20 (the number of initial jets) and taking into account the 
lower density of the atmosphere, where the centauros are produced, one ends up
with the estimate of about 10-100 events per km$^2$ per year, which is one to two 
orders of magnitude smaller than the claimed intensity of the Centauros \cite{ewa}. 
However, it seems to us that there is considerable uncertainty in the neutrino 
flux, which could take care of this discrepancy \cite{kalashev}.

$\bullet$ {\it Decay products}, $p_T , N_{\rm hadron}/N_{\rm em}$. The black hole, 
depending on its mass, decays initially to 10-20 fundamental particles of all 
kinds and with equal probability. Their energies are $\sim$100 GeV each in the 
black hole frame. The MBH emits almost as a black body of temperature 
$\sim$1 TeV, all types of matter and force quanta of the Standard Model.
The simplest possibility is that these initial partons, with $p_T$s also around
50-100 GeV will form hadrons, mostly mesons, of all kinds.  
The charmed or heavier mesons will decay to lighter ones, the neutral pions will 
decay to photons almost immediately, but the kaons will 
survive a distance of a few hundred meters. Since the Ks are counted as hadrons,
the ratio of $N_{\rm hadron}/N_{\rm em}$ will be enhanced, if the observation takes 
place at less than a few hundred meters from the initial interaction. 
The total multiplicities of the final showers in this case are a few decades \cite{cct}.

An alternative possibility is that a hot DCC forms before hadronization of the 
produced partons. This is known to lead to larger numbers of heavier mesons, which
makes it more probable to obtain a large ratio of the hadronic to the 
electromagnetic components. However, it is unclear if the system passes through 
such a DCC state. In any case, it seems that the probability to obtain a superclean 
Centauro event is rather small.

On the other hand, given the typically large values of $p_T$, 
the present scenario seems to be the most natural one to explain
this feature of the Centauros \cite{ewa2}.

$\bullet$ {\it Deep penetration}. The deep penetration follows from the assumption
of neutrinos, or some other weakly interacting particle (WIMP?), as the primary 
source of these events. According to the present picture any black holes produced
at higher altitudes from the detectors, will give signals similar to standard 
events \cite{cct}.  This kind of signature with large hadronic to electromagnetic
ratio, could only be obtained from the decay of black holes near the detector.

$\bullet$ {\it Halo}. At a more speculative level, it has been claimed 
recently \cite{cir} that after evaporation
a MBH is expected to leave behind a highly excited string state 
(a string ball), which in principle will decay to light (compared to $M_*$)
particles. Their life-time depends strongly on the excitation level and can be
considerably larger than $M_*^{-1}$. These objects may be serious candidates
in the context of the present scenario for the projectiles responsible for
the halos observed mostly in Chirons.

\section{Discussion}

On the basis of the above qualitative presentation and of the results
of the first numerical
steps taken in \cite{cct,cct2}, it seems that the above scenario may account
quite succesfully (a) for the energy threshold of all these events 
of order ${\cal O}$(TeV) in the center of mass frame, (b) the total multiplicities 
of a few dozen particles, dependent on the number of extra dimensions \cite{cct}, 
(c) the large values of $p_T$, and (d) the number and the rough heights of first 
interaction of the Centauro/Chiron events. It seems less succesfull in giving the right
$N_{\rm hadron}/N_{\rm em}$ (especially for the superclean events), even though
due to large statistical fluctuations, the numerical simulations \cite{cct} have not
been conclusive yet.  

Several alternative proposals have been put forward to explain these mysterious
events \cite{ewa}. An effort was made in \cite{mmt} and \cite{ewa2} to compare the 
various scenaria.
It should be clear however, that the study, for instance, of the production and 
evaporation processes of a MBH or of Strange Quark Matter 
fireballs, of the potential formation of a
high temperature quark-gluon plasma phase or of a DCC, of the hadronization of the 
100 GeV partons, all rely on unknown aspects of the physics involved.
Much more work is necessary and all possible realizations of a given scenario 
have to be investigated, before one can safely 
fit the observational data. The fundamental importance of the issues involved 
deserves every effort.

\end{document}